\documentclass[a4paper,11pt]{article}
\usepackage{pos}
\usepackage{wrapfig}

\title{Evidence of shadowing in inelastic nucleon-nucleon cross section}

\author{Kari J. Eskola}
\author{Ilkka Helenius}
\author*{Mikko Kuha}
\author{Hannu Paukkunen}

\affiliation{University of Jyvaskyla, Department of Physics, \\
  P.O. Box 35, FI-40014 University of Jyvaskyla, Finland}

\affiliation{Helsinki Institute of Physics,\\
P.O. Box 64, FI-00014 University of Helsinki, Finland}

\emailAdd{kari.eskola@jyu.fi}
\emailAdd{ilkka.m.helenius@jyu.fi}
\emailAdd{mikko.a.kuha@student.jyu.fi}
\emailAdd{hannu.t.paukkunen@jyu.fi}

\abstract{The Glauber modeling plays a key role in centrality-dependent measurements of heavy-ion collisions.
A central input parameter in Glauber models is the inelastic nucleon-nucleon cross section $\sigma_{\text{nn}}^{\text{inel}}$ which is nearly always taken from proton-proton measurements. At the LHC energies $\sigma_{\text{nn}}^{\text{inel}}$ depends on the QCD dynamics at small $x$ and low interaction scales where the shadowing/saturation phenomena are expected to become relatively more important for larger nuclei than for the proton. Thus, $\sigma_{\text{nn}}^{\text{inel}}$ e.g. in Pb+Pb collisions may well be lower than what is seen in proton-proton collisions. In this talk, we demonstrate how to use the recent $W^\pm$ and $Z$ measurements as a "standard candle" to extract $\sigma_{\text{nn}}^{\text{inel}}$ in Pb+Pb collisions. Our analysis -- built on the ATLAS data, state-of-the-art NNLO QCD calculations and nuclear PDFs -- indicate that at the LHC energies $\sigma_{\text{nn}}^{\text{inel}}$ in Pb+Pb collisions is suppressed relative to the proton-proton measurements by tens of percents. We demonstrate that this is in line with expectations from nuclear PDFs.}

\FullConference{%
  HardProbes2020\\
  1-6 June 2020\\
  Austin, Texas}

\begin{document}
\maketitle

\section{Introduction}

\begin{figure}[b]
\includegraphics[width=0.49\textwidth]{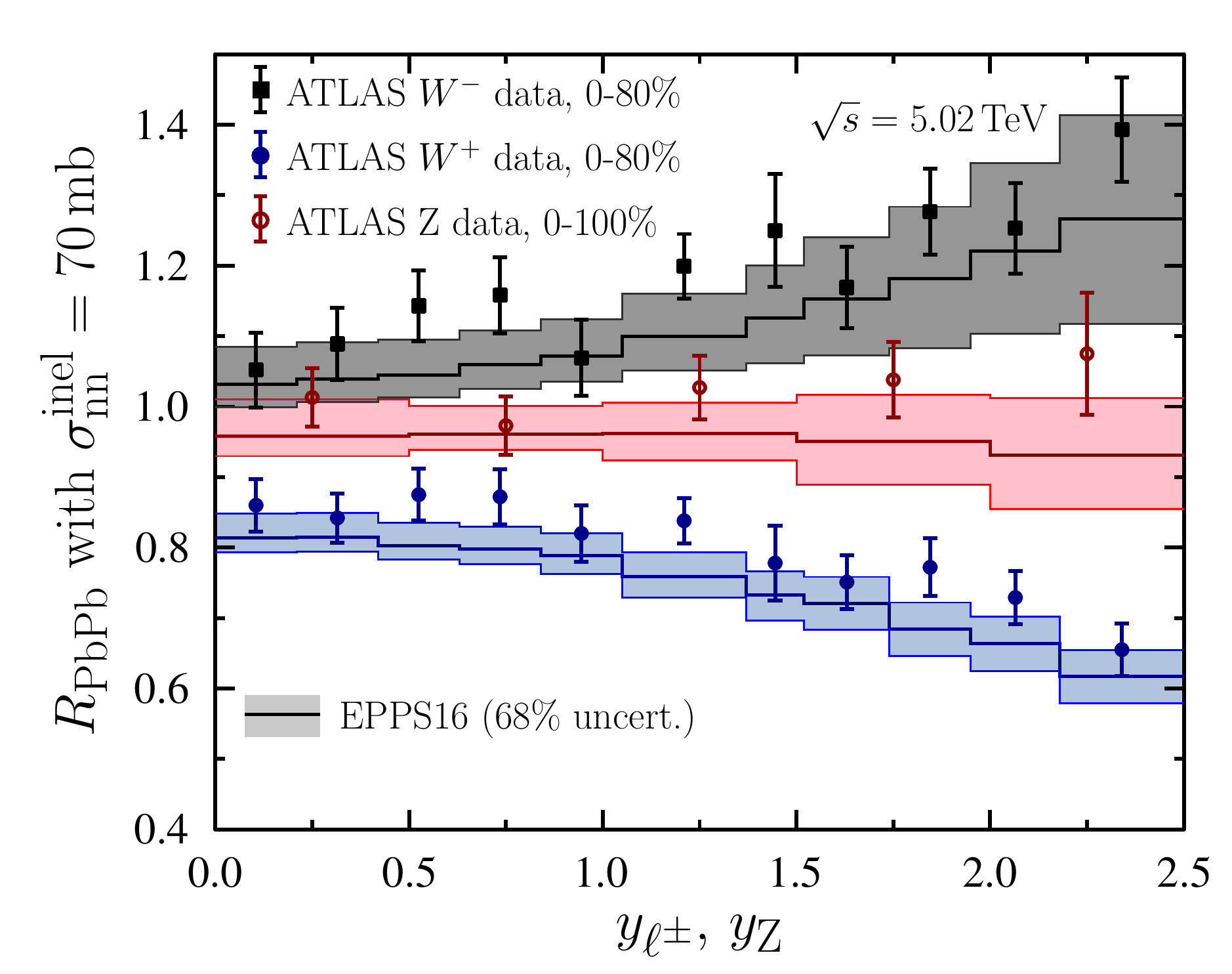}
\includegraphics[width=0.49\textwidth]{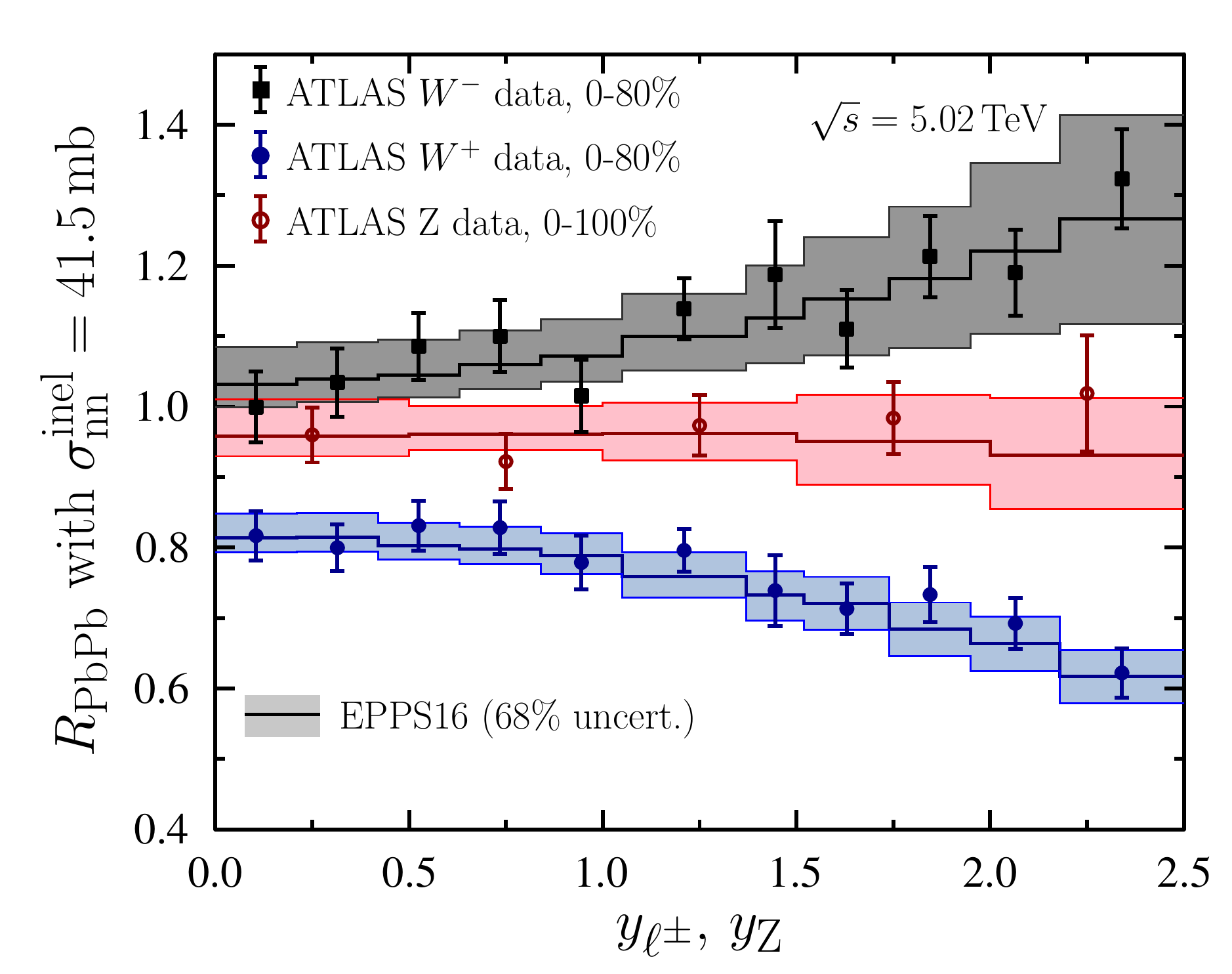}
\caption{The nuclear modification factor $R_{\mathrm{PbPb}}$ as a function of rapidity $y$. Solid curves are calculated using NNLO pQCD with NNPDF3.1 baseline proton PDFs and EPPS16 nuclear modifications. The error bands are the 68\% confidence intervals calculated from the nPDF error sets. In the left panel, data points are from \cite{ATLAS_W, ATLAS_Z}, with a normalization using $\sigma^{\textrm{inel}}_{\textrm{nn}}=70\;$mb. In the right panel, the same data points are re-normalized with $\langle T_{AA} \rangle$ calculated using nuclear-suppressed $\sigma^{\textrm{inel}}_{\textrm{nn}}=41.5\;$mb. Based on \cite{me}.}
\label{F:RAA_ATLAS_rap}
\end{figure}

Monte Carlo (MC) Glauber model is routinely used to e.g. turn per-event yields into centrality-dependent hard cross sections in the heavy-ion measurements at LHC and RHIC. Electroweak (EW) boson production in heavy-ion collisions is a “standard candle” that can be used to test factorization in nuclear collisions and, as argued in Ref. \cite{me}, also to study and calibrate Glauber model inputs. Here we have considered the recent high-precision analyses for $W^{\pm}$ and $Z$ boson production by ATLAS \cite{ATLAS_W,ATLAS_Z} based on Run II data for Pb+Pb collisions at $\sqrt{s_{\mathrm{nn}}} = 5.02~\text{GeV}$. We propose a new, data-driven, method to extract the nuclear-modified value for the inelastic nucleon-nucleon cross section.

\section{The nuclear modification factor $R_{\mathrm{PbPb}}$}

The main observable we studied was the centrality dependent nuclear modification factor $R_{\mathrm{PbPb}}$, which is experimentally defined as
\begin{equation}\label{E:RAAexp}
	R_{\mathrm{PbPb}}^{\mathrm{exp}}(y) = \frac{1}{\langle T_{AA} \rangle }\frac{\frac{1}{N_{\mathrm{evt}} }\mathrm{d}N^{W^{\pm}, Z}_{\mathrm{PbPb}}/\mathrm{d} y}{\mathrm{d}\sigma^{W^{\pm}, Z}_{\mathrm{pp}}/\mathrm{d}y},
\end{equation}
where the numerator is the per event yield, binned in rapidity $y$, and $\mathrm{d}\sigma^{W^{\pm}, Z}_{\mathrm{pp}}/\mathrm{d}y$ is the corresponding proton-proton cross section. The expectation value of nuclear overlap is calculated as $\langle T_{AA} \rangle  = \langle N_{\textrm{bin}} \rangle / \sigma^{\textrm{inel}}_{\textrm{nn}}$, where $\sigma^{\textrm{inel}}_{\textrm{nn}}$ is the nucleon-nucleon inelastic cross section. The expectation value of the number of colliding binary nucleon pairs $\langle N_{\textrm{bin}} \rangle$, which depends on centrality, is obtained from a MC Glauber model. As $\sigma^{\textrm{inel}}_{\textrm{nn}}$ is also a model parameter for MC Glauber, $\langle T_{AA} \rangle$ depends on it in a nontrivial way. By default, a nominal value of $\sigma^{\textrm{inel}}_{\textrm{nn}}(\sqrt{s_{\textrm{nn}}}=5.02\;\mathrm{TeV}) = 70\pm5\;$mb from fits to free proton-proton data is used.

\begin{figure}
    \vspace{-0.35em}
    \includegraphics[width=0.49\textwidth]{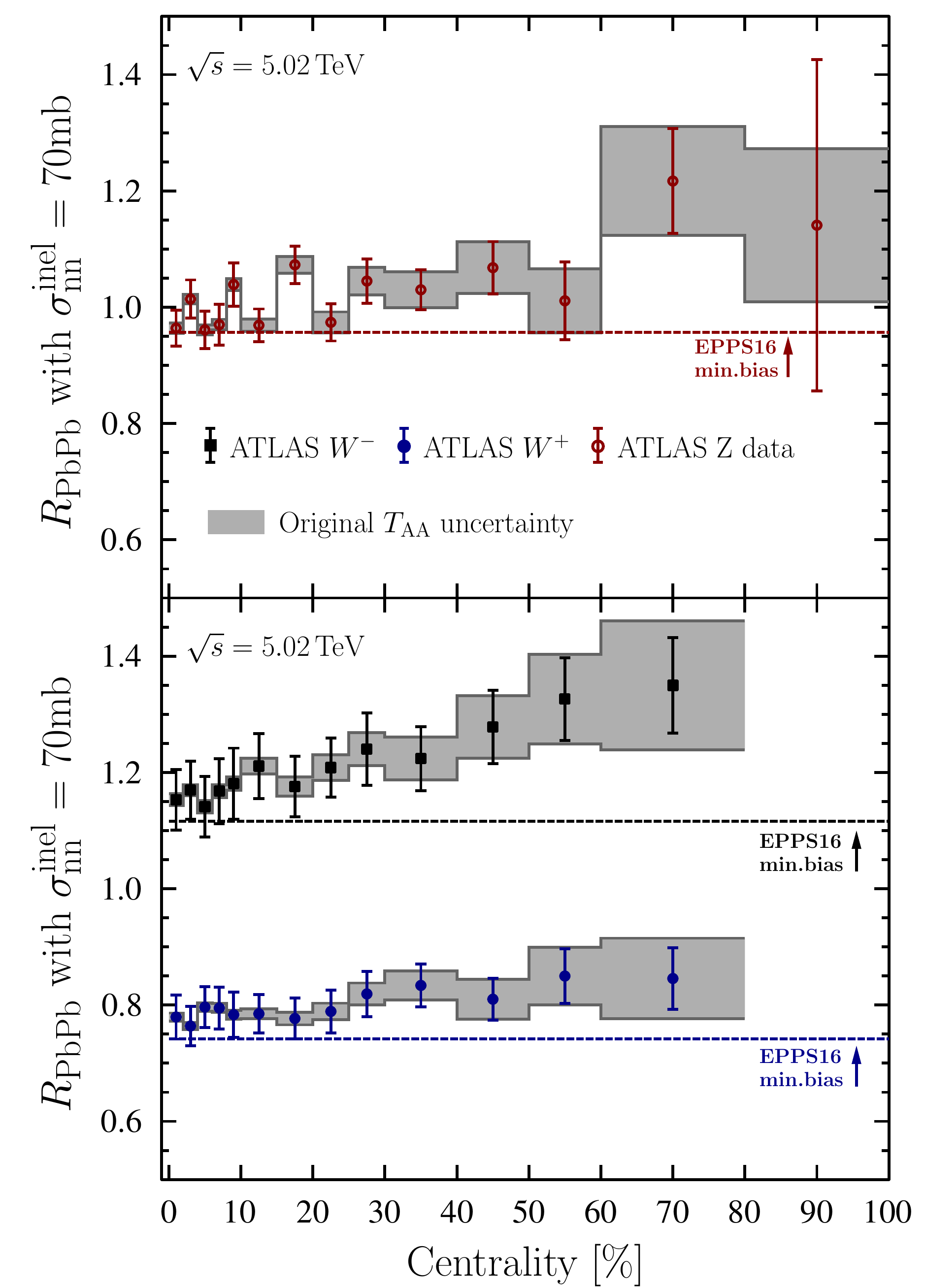}
    \includegraphics[width=0.49\textwidth]{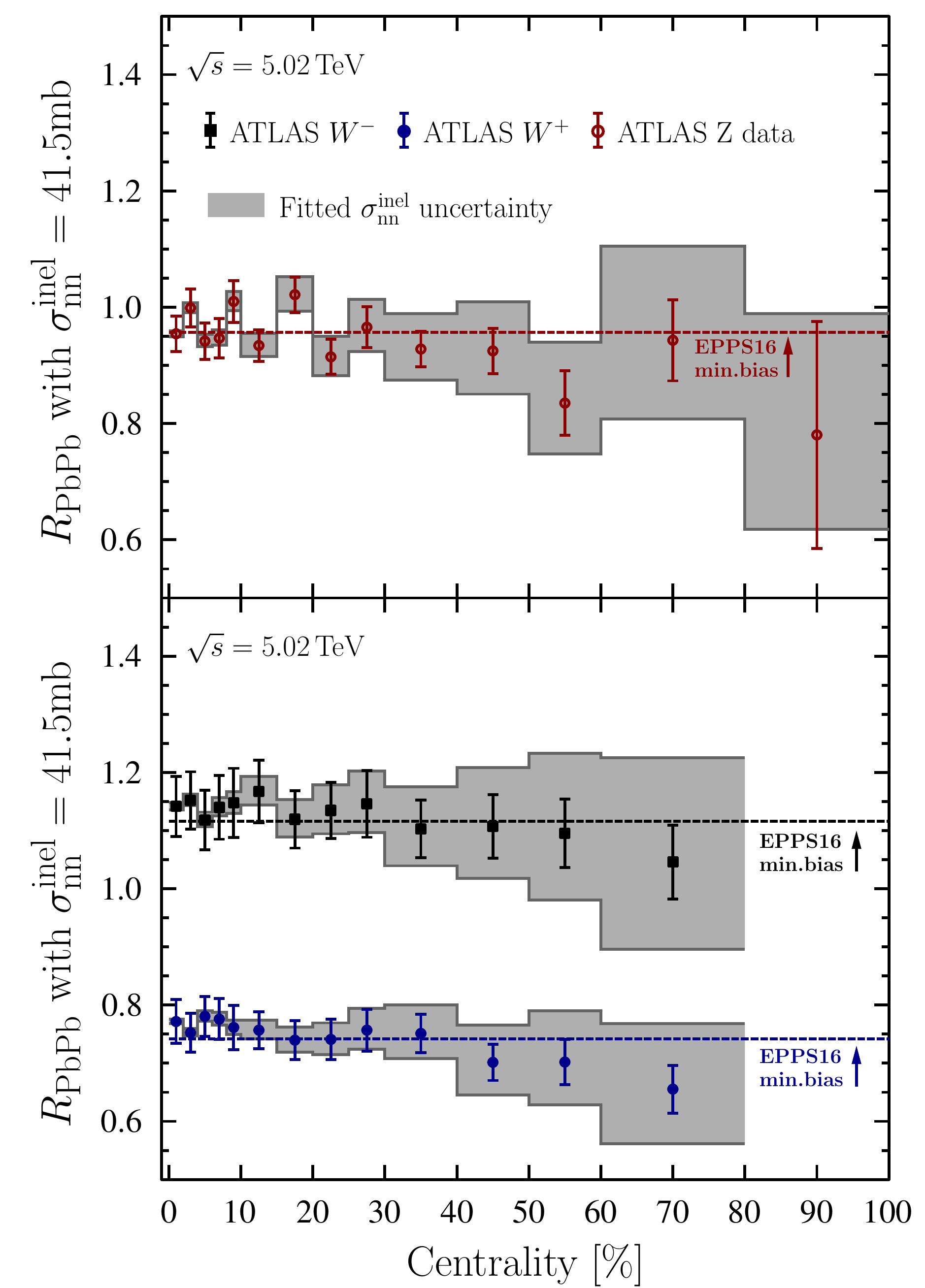}
    \caption{The nuclear modification factor $R_{\mathrm{PbPb}}$ as a function of centrality. In the left panel, data points are from \cite{ATLAS_W,ATLAS_Z}, with a normalization using $\sigma^{\textrm{inel}}_{\textrm{nn}}=70\;$mb. In the right panel, the same data points are re-normalized with $\langle T_{AA} \rangle$ calculated using nuclear-suppressed $\sigma^{\textrm{inel}}_{\textrm{nn}}=41.5\;$mb. From \cite{me}.}
    \label{F:RAA_cent_combo}
\end{figure}

The centrality-integrated nuclear modification factor can be theoretically defined as 
\begin{equation}\label{E:RAAtheor}
	R_{\mathrm{PbPb}}^{\mathrm{theor}}(y) = \frac{1}{(208)^2} \frac{\mathrm{d}\sigma^{W^{\pm}, Z}_{\mathrm{PbPb}}/\mathrm{d} y}{\mathrm{d}\sigma^{W^{\pm}, Z}_{\mathrm{pp}}/\mathrm{d}y},
\end{equation}
where $\sigma^{W^{\pm}, Z}_{\mathrm{PbPb}}$ is an inclusive EW cross section, where the nuclear effects are encoded into nuclear parton distribution functions (nPDFs). In this analysis, the cross sections were calculated at the next-to-next-to-leading order (NNLO) in perturbative QCD (pQCD) using MCFM \cite{MCFM}. We used NNPDF3.1 \cite{NNPDF} as the proton PDFs, as they match the ATLAS results for EW boson production in p+p at $\sqrt{s}=5.02\;$TeV very well \cite{ppdata}. For nuclear modifcations, we used EPPS16 \cite{EPPS}, as they provide a good description of the $W^{\pm}$ production data in p+Pb collisions \cite{pPbdata}. 

The comparison of calculated and measured $R_{\mathrm{PbPb}}$ is shown in the left panel of Figure \ref{F:RAA_ATLAS_rap}. The $y$-dependence seems to be correctly described with the EPPS16 nPDFs but there is a clear hint of a normalization difference as the data lie slightly above the calculated result. This suggests that we could try to use these data to calibrate the Glauber model. Similarly in the left panel of Figure \ref{F:RAA_cent_combo}, there is a peculiar trend of rising $R_{\mathrm{PbPb}}$ towards the more peripheral collisions. One might have expected a flat behaviour or, if anything, a slight downward trend due to geometrical and selection biases associated with MC Glauber modelling \cite{RAA_biases}. It is worth noting that the centrality classes of the $W^\pm$ data in Figure \ref{F:RAA_ATLAS_rap} are different from those of the $Z$ data, but this has only a negligible effect as the most peripheral centrality bins carry a very insignificant weight in the integrated observable.

\section{The data-driven value of $\sigma^{\textrm{inel}}_{\textrm{nn}}$}

\begin{figure}
	\begin{minipage}[b]{0.48\textwidth}
        \includegraphics[width=\textwidth]{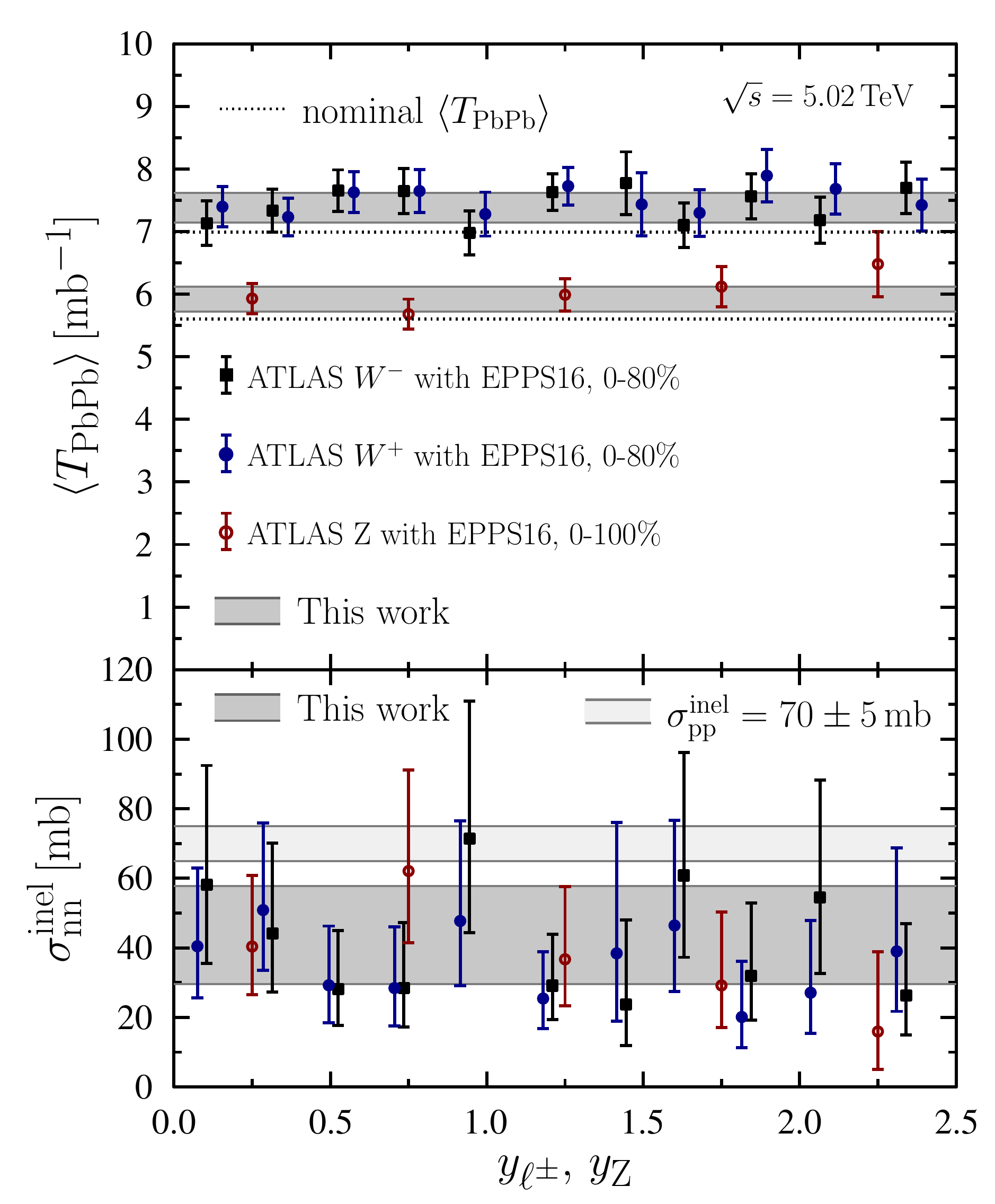}
        \caption{$\langle T_{AA} \rangle$ and $\sigma^{\textrm{inel}}_{\textrm{nn}}$ preferred by the EW data as a function of rapidity $y$. The darker grey bands are the obtained best fit. The light grey band and the dotted lines are the nominal values. From \cite{me}.}
        \label{F:fit_result}
    \end{minipage}\hfill
	\begin{minipage}[b]{0.48\textwidth}
        \includegraphics[width=\textwidth]{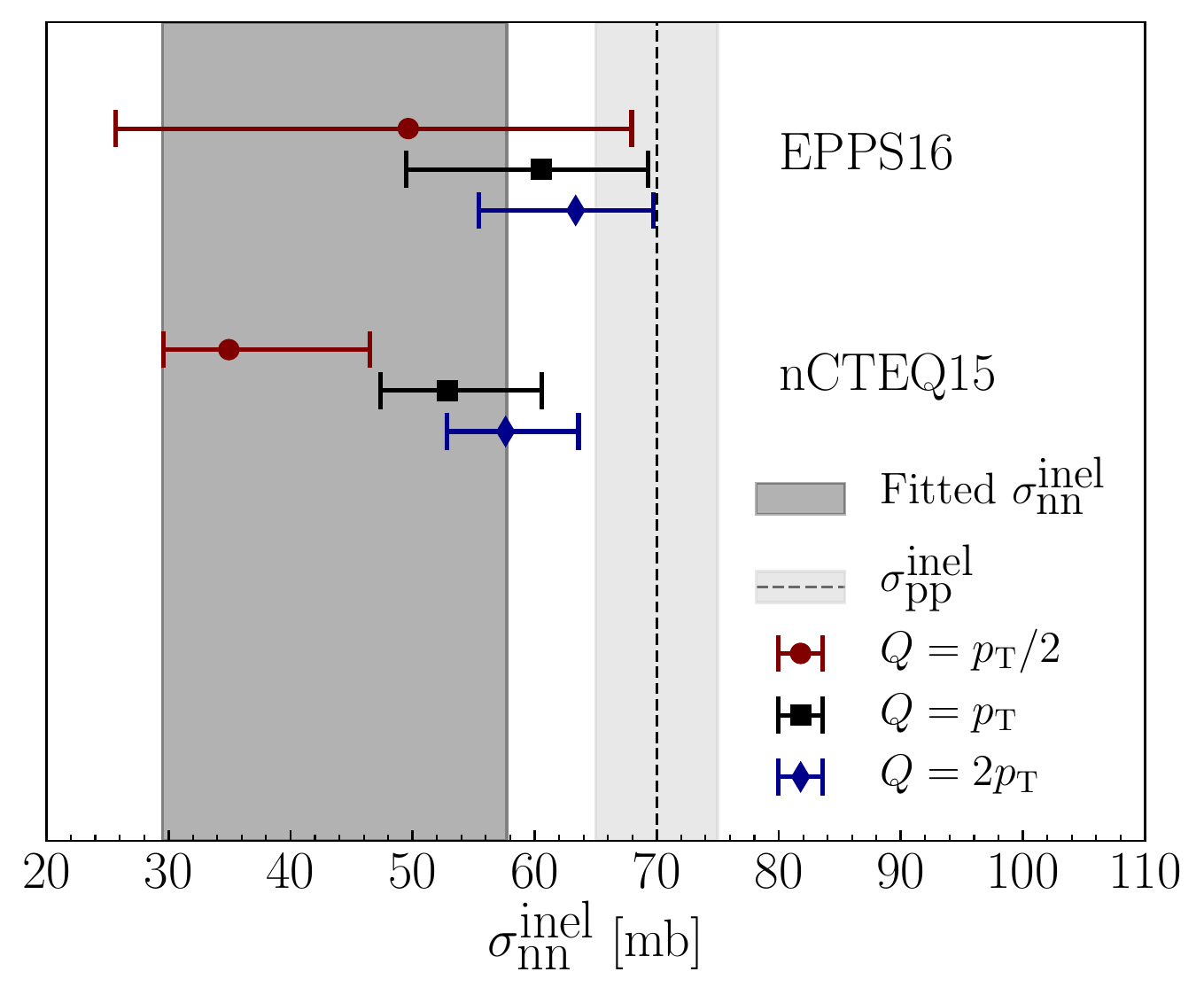}
        \caption{ $\sigma^{\textrm{inel}}_{\textrm{nn}}$ calculated in an eikonal minijet model using EPPS16 and nCTEQ15 nuclear shadowing, with three different factorization/renormalization scale $Q$ choices. The error bars represent the 68\% confidence limit uncertainties of the nPDF sets.  The dark grey band is our best fit result of $\sigma^{\textrm{inel}}_{\textrm{nn}}=41.5^{+16.2}_{-12.0}\;$mb, and the light grey band is the nominal $\sigma^{\textrm{inel}}_{\textrm{pp}}=70\pm5\;$mb. From \cite{me}.}
        \label{F:minijets}
    \end{minipage}
\end{figure}

The MC Glauber model maps $\sigma^{\textrm{inel}}_{\textrm{nn}}$ bijectively into $\langle T_{AA} \rangle$ for each centrality class. Therefore, to find the normalization preferred by the data, we can try to vary $\sigma^{\textrm{inel}}_{\textrm{nn}}$, and by that $\langle T_{AA} \rangle$, so that re-normalized $R_{\mathrm{PbPb}}^{\mathrm{exp}}$ from \eqref{E:RAAexp} matches $R_{\mathrm{PbPb}}^{\mathrm{theor}}$ from \eqref{E:RAAtheor}. In practice, we did this by minimizing
\begin{equation}\label{E:chisqr}
	\textstyle\chi^2 =  \sum_{i} \left[ \scriptstyle\frac{R_i^{\mathrm{exp}}\times \left. \frac{\langle T_{AA}(\sigma_{\mathrm{pp}}^{\mathrm{inel}})\rangle}{\langle T_{AA}	(\sigma^{\textrm{inel}}_{\textrm{nn}})\rangle} \right. -  R_i^{\mathrm{theor}} + \sum_k f_k \beta_i^k}{\delta_i^{\mathrm{exp}} \times \left. \frac{\langle T_{AA}(\sigma_{\mathrm{pp}}^{\mathrm{inel}})\rangle}{\langle T_{AA}(\sigma^{\textrm{inel}}_{\textrm{nn}})\rangle} \right.} \textstyle \right]^2  +\, T \sum_k f_k^2 , 
	\quad\quad
	\beta_i^k \equiv \frac{1}{2}\left[  R_i^{\mathrm{theor}}(S_k^+ ) -  R_i^{\mathrm{theor}}(S_k^- )\right],
\end{equation}
with respect to $\sigma^{\textrm{inel}}_{\textrm{nn}}$ and $f_k$:s. The index $i$ labels the separate data points and $k$ the EPPS16 nPDF error set pairs. $\delta_i^{\mathrm{exp}}$ is the experimental uncertainty of a data point and $S_k^\pm$ are the nPDF error sets. The factor $T=1.645^2$ in the penalty term scales the uncertainty interval to the 68\% confidence level. This procedure gave us a best fit value of 
\begin{equation*}
	\sigma^{\textrm{inel}}_{\textrm{nn}}=41.5^{+16.2}_{-12.0}\;\textrm{mb},
\end{equation*}
where the uncertainty is given by the $\Delta\chi^2=1 $ criterion. The result of the fit is depicted in Figure \ref{F:fit_result}. The data points are obtained by equating \eqref{E:RAAexp} and \eqref{E:RAAtheor}, and solving for $\langle T_{AA} \rangle$, separately for each $R_{\mathrm{PbPb}}^{\mathrm{exp}}(y)$ point in \cite{ATLAS_W,ATLAS_Z}. MC Glauber maps $\sigma^{\textrm{inel}}_{\textrm{nn}}$ to $\langle T_{AA} \rangle$ for each centrality class, so there is a one-to-one correspondence between the two panels.

When this, significantly suppressed value of $\sigma^{\textrm{inel}}_{\textrm{nn}}$ is used in the normalization of the data (seen in the right panel of Figure \ref{F:RAA_ATLAS_rap}), one finds a very good agreement between the theoretically calculated $R_{\mathrm{PbPb}}$ and the experimental result. Also, due to the most peripheral centrality classes of $\langle T_{AA} \rangle$ being the ones affected the most by the modification of $\sigma^{\textrm{inel}}_{\textrm{nn}}$, the mysterious increasing trend in $R_{\mathrm{PbPb}}$ as a function of centrality vanishes (seen in the right panel of Figure \ref{F:RAA_cent_combo}).

\section{Nuclear suppression of $\sigma^{\textrm{inel}}_{\textrm{nn}}$ from eikonal minijet model}

To find a justification for the large suppression of  $\sigma^{\textrm{inel}}_{\textrm{nn}}$, we studied an eikonal model for minijet production including nuclear shadowing. The model we used is similar to the one in \cite{XNW_minijets}, but it includes no soft component in the eikonal function nor a $K$-factor, but only the hard minijet cross section $\sigma_{\textrm{jet}}(\sqrt{s_{nn}}, p_0)$ to the lowest order of pQCD. We use the transverse momentum cutoff $p_0$ and partonic overlap width as parameters to fit the model to experimental data on inelastic and total p+p cross sections. After fixing the model in p+p, we turn on the nuclear modifications in the PDFs to estimate the magnitude of the effect of nuclear shadowing in Pb+Pb. For the free proton, we used CT14LO PDFs, and for the nuclear modifications we studied both EPPS16 \cite{EPPS} and nCTEQ15 \cite{nCTEQ}. As seen in Figure \ref{F:minijets}, we found a similar amount of suppression for $\sigma_{\mathrm{nn}}^{\mathrm{inel}}$ (several tens of percents) from this simplistic model with both of the nPDF sets, as we did from our fit to the ATLAS data.

\section{Summary}

We compared the state-of-the-art pQCD calculation to the measured $W^\pm$ and $Z$ boson nuclear modification factor $R^{W^{\pm}, Z}_{\mathrm{PbPb}}$ to obtain the nuclear-suppressed value for $\sigma^{\textrm{inel}}_{\textrm{nn}}$ at $\sqrt{s_{\mathrm{nn}}} = 5.02~\text{TeV}$. The recent high-precision ATLAS EW data from LHC run II prefer a significantly suppressed $\sigma^{\textrm{inel}}_{\textrm{nn}}=41.5^{+16.2}_{-12.0}\;\textrm{mb}$, down from the nominal $70\pm5\;$mb. This amount of suppression is consistent with nuclear shadowing of PDFs in the eikonal minijet model calculation. Such a suppression in $\sigma_{\mathrm{nn}}^{\mathrm{inel}}$ would affect all experimental analyses relying on Glauber models in obtaining the normalization factors for different centrality classes.

\acknowledgments

We thank the Academy of Finland, projects 297058 (K. J. E.) and
308301 (H. P. and I. H.), and the Väisälä Foundation
(M. K.) for financial support. Computing resources from CSC – IT Center for Science in Espoo, Finland were used (project jyy2580).


\begin{thebibliography}{99}

\bibitem{me}
K. J. Eskola, I. Helenius, M. Kuha, and H. Paukkunen,
\href{https://arxiv.org/abs/2003.11856}{arXiv:2003.11856 [{\tt hep-ph}]}.\\[-2.5em]

\bibitem{ATLAS_W}
G. Aad \textit{et al.}  (ATLAS collaboration), 
\href{https://dx.doi.org/10.1140/epjc/s10052-019-7439-3}{\emph{Eur. Phys. J.} \textbf{C79} (2019) 935}.\\[-2.5em]

\bibitem{ATLAS_Z}
G. Aad \textit{et al.}  (ATLAS collaboration), 
\href{https://dx.doi.org/10.1016/j.physletb.2020.135262}{\emph{Phys. Lett.} \textbf{B802} (2020) 135262}.\\[-2.5em]

\bibitem{MCFM}
 R. Boughezal \textit{et al.}, 
\href{https://dx.doi.org/10.1140/epjc/s10052-016-4558-y}{\emph{Eur. Phys. J.} \textbf{C77} (2017) 7}.\\[-2.5em]

\bibitem{NNPDF}
R. D. Ball \textit{et al.}  (NNPDF collaboration), 
\href{https://dx.doi.org/10.1140/epjc/s10052-017-5199-5}{\emph{Eur. Phys. J.} \textbf{C77} (2017) 663}.\\[-2.5em]

\bibitem{ppdata}
M. Aaboud \textit{et al.}  (ATLAS collaboration), 
\href{https://dx.doi.org/10.1140/epjc/s10052-019-6622-x}{\emph{Eur. Phys. J.} \textbf{C79} (2019) 128}.\\[-2.5em]

\bibitem{EPPS}
K. J. Eskola, P. Paakkinen, H. Paukkunen, and C. A. Salgado, 
\href{https://dx.doi.org/10.1140/epjc/s10052-017-4725-9}{\emph{Eur. Phys. J.} \textbf{C77} (2017) 163}.\\[-2.5em]

\bibitem{pPbdata}
A. M. Sirunyan \textit{et al.}  (CMS collaboration), 
\href{https://dx.doi.org/10.1016/j.physletb.2019.135048}{\emph{Phys. Lett.} \textbf{B800} (2020) 135048}.\\[-2.5em]

\bibitem{RAA_biases}
C. Loizides and A. Morsch, 
\href{https://dx.doi.org/10.1016/j.physletb.2017.09.002}{\emph{Phys. Lett.} \textbf{B773} (2017) 408}.\\[-2.5em]

\bibitem{XNW_minijets}
X.-N. Wang, 
\href{https://doi.org/10.1103/PhysRevD.43.104}{\emph{Phys. Rev.} \textbf{D43} (1991) 104}.\\[-2.5em]

\bibitem{nCTEQ}
K. Kovarik \textit{et al.},
\href{https://dx.doi.org/10.1103/PhysRevD.93.085037}{\emph{Phys. Rev.} \textbf{D93} (2016) 085037}.\\[-2.5em]

\end{thebibliography}
\end{document}